# A Hierarchical Modulation for Upgrading Digital Broadcast Systems

Hong Jiang and Paul Wilford
Bell Laboratories, Lucent Technologies Inc.
700 Mountain Ave, P.O. Box 636, Murray Hill, NJ 07974-0636

*Abstract* — **A hierarchical modulation scheme is proposed to upgrade an existing digital broadcast system, such as satellite TV, or satellite radio, by adding more data in its transmission. The hierarchical modulation consists of a basic constellation, which is the same as in the original system, and a secondary constellation, which carries the additional data for the upgraded system. The upgraded system with the hierarchical modulation is backward compatible in the sense that receivers that have been deployed in the original system can continue receiving data in the basic constellation. New receivers can be designed to receive data carried in the secondary constellation, as well as those in the basic constellation. Analysis will be performed to show the tradeoff between bit rate of the data in secondary constellation and the penalty to the performance of receiving the basic constellation.**

## I. INTRODUCTION

Digital broadcast systems have increasingly been deployed for services such as terrestrial digital TV, digital radio, satellite TV and radio, and digital cable TV. A digital broadcast system utilizes regulated frequency bands with fixed bandwidths. The capacity of a digital broadcast system is limited by transmission power of the system and channel impairments. Since in a broadcast system, the same data is transmitted to all users, there is a tradeoff between transmitted bit rates and intended coverage areas. A digital broadcast system is usually designed with a bit rate that can be reliably received by users in an intended coverage area for a given transmission power.

Some digital broadcast systems are designed with a flexibility of transmitting with different bit rates by allowing different modulating constellations (such as QPSK, 8PSK, 16QAM, 64QAM etc), and error correction codes of different coding rates. It is desirable for receivers in such a system to be capable of receiving data at all of the specified bit rates. An advantage of such a system is that it is easy, at least in concept, to upgrade a deployed system to increase the bit rate, simply by switching to a modulation with a larger constellation and/or an error correction code of higher rate. A disadvantage of such an arrangement is that all receivers have to be designed with capability of receiving all specified rates, and hence such receivers may be expensive to manufacture.

Many deployed systems, however, do not have such flexibility; they are designed with a fixed modulation scheme, and error correction codes of fixed rates. Examples of such systems are the original ATSC terrestrial 8VSB system, and some digital satellite radio and TV systems. Oftentimes, after such a system is deployed and in service for a period of time, there are rising needs to upgrade the system to provide more services than the system was originally designed. The needs for upgrading the system may arise for many reasons. There is a market pressure for a service provider to add more services in order to differentiate itself from its competitors. New services may also generate additional revenue. On the technical side, new algorithms may have been developed to allow higher bit rate to be transmitted through the same channel, and at the same time, advances in technology make the implementation of the new algorithms feasible and cost-effective. This advance in algorithm development and technology makes it possible to build a new generation of receivers that can have much better performance under the same channel conditions. For example, the utilization of iterative decoding algorithms such as BCJR [1, 2, 3] and the MIMO technology [4] can provide additional gains over receivers using traditional algorithms, or a single antenna.

Backward compatibility is a major difficulty in increasing the bit rate of a fixed rate digital broadcast system that has been already deployed for a period of time. Since the system has already been deployed, there are a large number of receivers already used by subscribers that can only receive signals from the original system with the specific modulation and bit rate. To replace these deployed receivers by new generation of receivers may be prohibitively expensive. Therefore, any upgrade of the system must be backward compatible, in a way that the deployed receivers must continue to operate in the upgraded system, even though the deployed receivers may not be able to receive the supplementary data that is added to the original system.

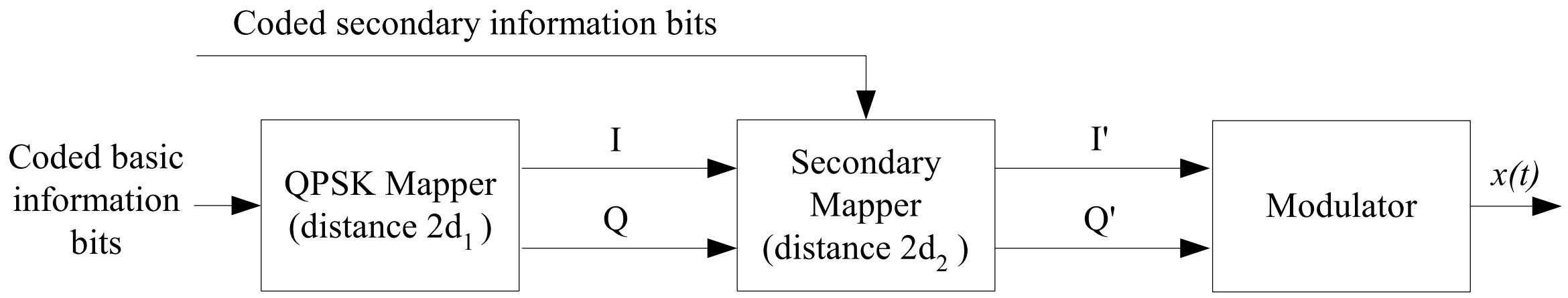

**Figure 1** QPSK/16QAM Modulation

In this paper, we propose a hierarchical modulation [5, 6, 10] method to increase the bit rate of a deployed digital broadcast system. The upgraded system is backward compatible, so that the upgrade is transparent to the deployed receivers of the original system. The hierarchical modulation has already been included in the DVB-T standard [12].

We will use the term *basic information* to mean data transmitted in the originally designed system, as is customary in the literature of hierarchical modulations. The data that is added in the upgraded system will be named *secondary information*. Thus, the originally designed system with deployed receivers transmits basic information only, while the upgraded system transmits both basic and secondary information.

Although the concept of this paper may be applied to hierarchical modulation of any constellations, we will limit our discussions to QPSK/16QAM hierarchical modulation in the remainder of this paper. The reason is twofold. First, most digital satellite systems use QPSK, and due to its small constellation size, a QPSK system not only has the most urgent need for, but also can benefit the most from the addition of a secondary information channel. Secondly, the use of a specific hierarchical constellation, QPSK/16QAM in this case, simplifies the analysis.

The remainder of this paper is organized as follows. In the next section, the hierarchical modulation for the basic and secondary information is described. An analysis of the BER in terms of CNR and the ratio of minimum distances is given in Section III. Also, in Section III, penalties of the hierarchical modulation to the originally designed receivers are derived. In Section IV, a design of new generation receivers is presented for the hierarchical modulation in which an iterative decoding algorithm is used to provide additional channel gains for both basic and secondary information bits. The bit rate at which the secondary information can be reliably received is also estimated. Finally in Section V, we address the issue of the local information transmission.

## II. QPSK/16QAM HIERARCHICAL MODULATION

Hierarchical modulations were initially proposed to provide different classes of data to users with different reception conditions [5, 6, 10]. In that application, all users within the coverage range can reliably receive basic information, while users under more favorable conditions can receive additional "refinement information".

In the context of the current application of upgrading a deployed QPSK system, the hierarchical QPSK/16QAM modulation can be described by the block diagram in Figure 1. The basic and secondary information bits are channel encoded. The coded basic information bits are mapped to the QPSK constellation in the same way as the original QPSK system. The minimum distance between points in the QPSK constellation is denoted by $2d_1$ . This represents the basic hierarchical constellation. The basic hierarchical constellation is next modified according to the coded secondary information bits, and the combined hierarchical constellation is formed as shown in Figure 2(b). The combined constellation is a 16QAM constellation with the minimum distance between two points denoted by $2d_2$ . The mapping of the basic and secondary information bits indicated in Figure 2(b) is a Karnaugh map style Gray mapping. In Figure 2(b), the two most significant bits (in bold face) represent the basic information bits while the two remaining bits represent the secondary information bits.

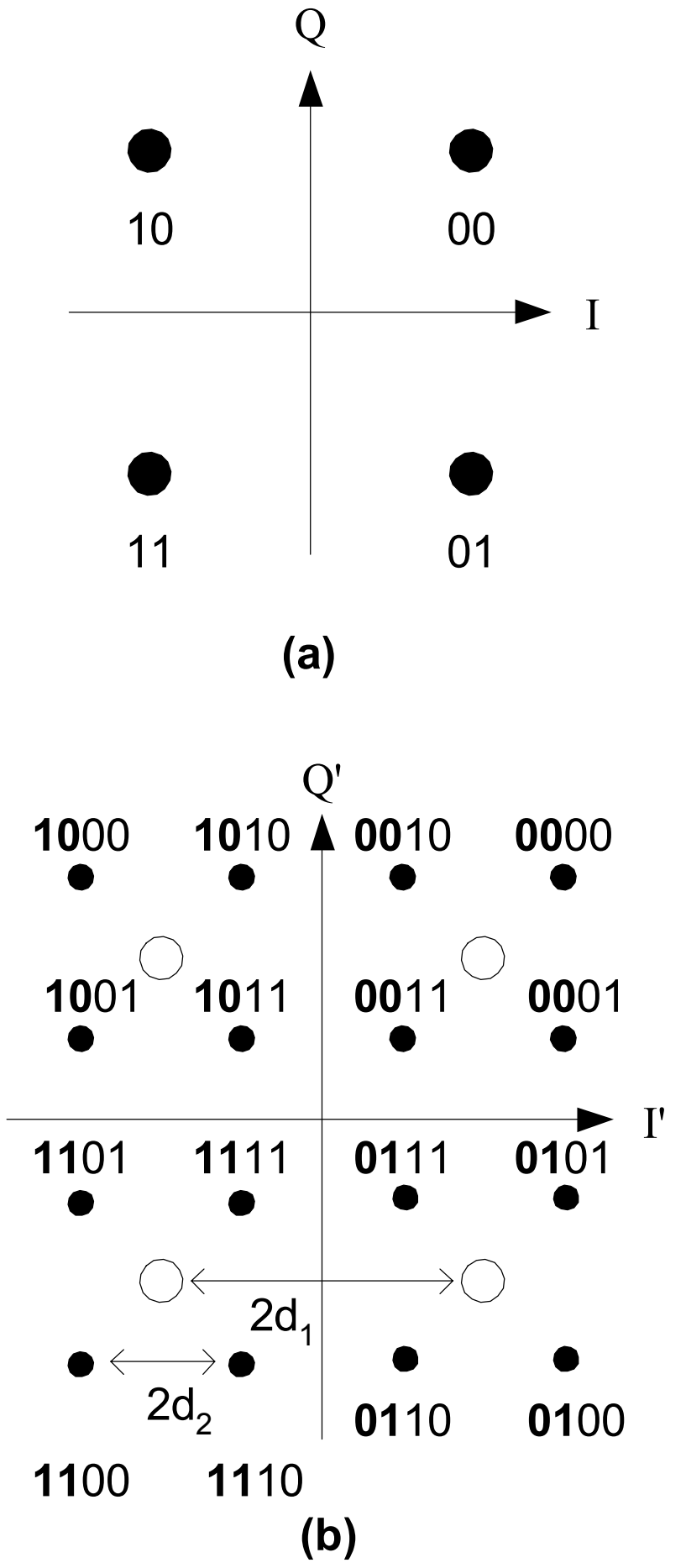

**Figure 2** QPSK/16QAM Hierarchical Constellation
**(a)** QPSK constellation
**(b)** 16QAM constellation. The bits in bold face are basic information bits, and the rest are secondary information bits

The two blocks, QPSK Mapper and Secondary Mapper, in Figure 1 may be combined into one. They are presented separately because the two blocks clearly demonstrate the concept of the hierarchical modulation used for upgrading an already deployed system: the secondary information bits are added to the transmission by modifying the constellation of the basic information.

Let $\lambda$ be the ratio of the minimum distances in the QPSK and 16QAM,

$$\lambda = \frac{d_2}{d_1}. \quad (1)$$

$\lambda$ is an important parameter to characterize the system. If $\lambda = 0$, it is the QPSK system, transmitting only the basic information. If $\lambda = \frac{1}{2}$, it is the uniform 16QAM, in which the basic and secondary information bits have same channel conditions. We will be only interested in the case with $0 < \lambda < \frac{1}{2}$, where the constellation is truly hierarchical.

When $\lambda$ is small, the four points in each quadrant of the constellation form a "cloud". To the originally designed receivers, a cloud represents a point in the QPSK constellation. For example, any point in the first quadrant, (the ones labeled **00**00, **00**01, **00**11, and **00**10 in Figure (b)) is treated as the point 00 in the QPSK constellation. The variation among the points in a cloud has the same effect of white noise on the originally designed receivers. Therefore, in the upgraded system after the secondary information is added, the originally designed receivers will continue to operate and receive the basic information bits, with the only difference being that they may operate at a higher noise level.

The additional noise due to the secondary information in the upgraded system imposes a penalty on the performance of originally designed receivers. We will analyze the penalty in terms of the hierarchy parameter $\lambda$ in the next section.

New receivers can be designed to operate in the hierarchical system. New receivers will be able to distinguish points in the "cloud", and extract both the basic and secondary information bits. We will address the issue of new receiver designs in section 4.

For convenience, we define the following terminology:

*QPSK systems* – systems with QPSK modulation. They refer to the systems before upgraded to the hierarchical modulation.

*QPSK receivers* – the originally designed receivers that are only capable of receiving the QPSK modulation or the basic information in the hierarchical modulation.

*Hierarchical systems* – systems with hierarchical modulation, referring to the upgraded systems with both basic and secondary information.

*Hierarchical receivers* – the new receivers that are designed to operate in the QPSK/16QAM system, and are capable of receiving both the basic and secondary information bits.

## III. PENALTY ANALYSIS

The carrier to noise ratio (CNR) of the hierarchical constellation of Figure 2(b) is given by

$$CNR = \frac{E_s}{N_0} = \frac{2(1+\lambda^2)d_1^2}{N_0}, \quad (2)$$

where $E_s$ is the carrier power and $N_0$ is the channel noise power.

When signals with the hierarchical constellation of Figure 2(b) are received by the QPSK receivers, the constellation is treated as QPSK constellation, with power $2d_1^2$. To these receivers, the noise consists of two terms, the channel noise $N_0$, and the scattering of points in the secondary hierarchy constellation, $2\lambda^2 d_1^2$. We define the modulation noise ratio (MNR) to be the ratio of the power of QPSK constellation to the combined noise power, and it is given by

$$\begin{aligned} MNR &= \frac{2d_1^2}{N_0 + 2\lambda^2 d_1^2} \\ &= \frac{1}{1+\lambda^2(1+CNR)} CNR. \end{aligned} \quad (3)$$

MNR can be considered as the CNR of the hierarchical constellation of Figure 2(b) as viewed by the QPSK receivers. It is a measure of the "cleanness" of the QPSK constellation as viewed by the QPSK receivers. The performance of these receivers, such as timing recovery and carrier recovery, may be characterized by *MNR*. Since the performance of these receivers is characterized by *CNR* in the QPSK system, the performances of these receivers before and after the secondary information is added may be evaluated by comparing the values of *CNR* and *MNR*. The difference between *CNR* and *MNR* is the penalty to the QPSK receivers. From equation (3), the penalty in MNR can be defined as

$$P_{MNR} = 1 + \lambda^2(1+CNR), \quad (4)$$

which is the ratio $\frac{CNR}{MNR}$.

The penalty, a function of both $\lambda$ and *CNR*, represents the additional carrier power that is needed in the hierarchical system so that the QPSK receivers can see the same cleanness of the constellation as in the QPSK system. It is a measure of how much the QPSK receivers suffer due to the addition of the secondary information. The larger the penalty is, the worse these receivers will perform in the hierarchical system.

A plot of $P_{MNR}$ vs. *CNR* for different values of $\lambda$ is shown in Figure 3. In most deployed satellite broadcast systems such as satellite radio, or satellite TV, the minimum operating CNR is below 7dB. That is, receivers can perform reliably at CNR higher than 7dB. As shown in Figure 3, at *CNR* = 7dB, the penalty is around 0.25dB for $\lambda = 0.1$, and $P_{MNR} \leq 0.5$ dB for $\lambda = 0.15$. This means that for $\lambda = 0.1$, when the transmission power in the hierarchical system has a 7dB CNR, the QPSK receivers effectively get a QPSK constellation with equivalence of *CNR* = 6.75dB, because the penalty is 0.25dB. For $\lambda = 0.15$, the QPSK receivers can receive a QPSK constellation equivalent to *CNR* = 6.5dB when the hierarchical constellation has *CNR* = 7dB. The penalty is even smaller at lower CNR. Although the penalty is higher at higher CNR, the penalty becomes insignificant because there are enough margins in the CNR to meet the desired performance. These results show that the penalty to the QPSK receivers is acceptable for $\lambda \leq 0.15$ under most operating conditions.

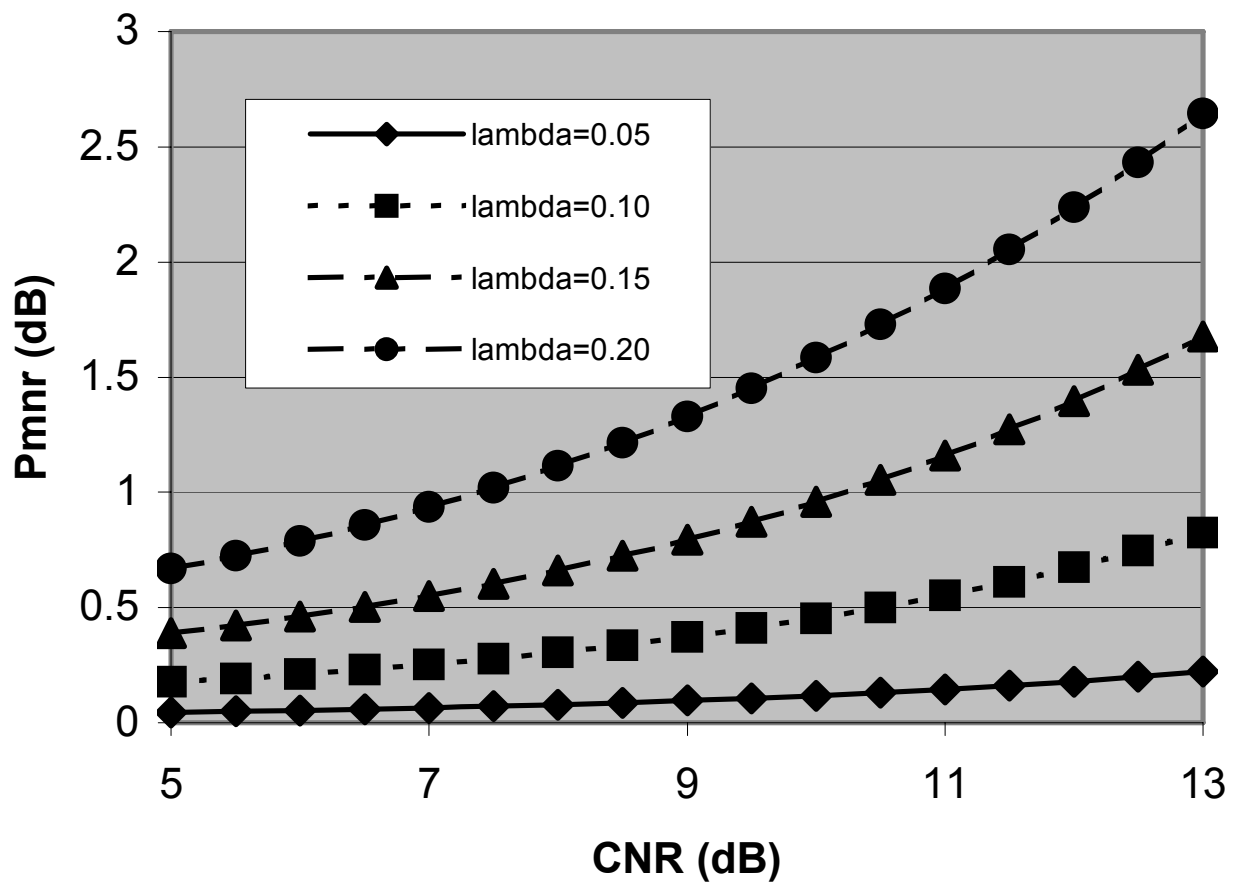

**Figure 3** Penalty in MNR

While MNR may be useful in evaluating performance of timing recovery and carrier recovery, it is not accurate for bit error rate (BER) estimate. In the following, we will use another approach to estimate BER of the QPSK receivers in the hierarchical system over additive white Gaussian noise (AWGN) channels.

BER computations of QPSK/16QAM over AWGN channels have been performed by Vitthaladevuni et al. [7]. The probability of raw bit error (without error correction coding) made by the QPSK receivers in the hierarchical system is given by

$$BER_B = \frac{1}{2}Q\left(\frac{1-\lambda}{\sqrt{1+\lambda^2}}\sqrt{CNR}\right) + \frac{1}{2}Q\left(\frac{1+\lambda}{\sqrt{1+\lambda^2}}\sqrt{CNR}\right), \quad (5)$$

where $Q(x) = \frac{1}{2}\text{erfc}\left(\frac{x}{\sqrt{2}}\right)$, and function erfc($x$) is the complementary error function. In the QPSK system before the secondary information bits are added, the QPSK constellation is transmitted and the probability of bit error is

$$BER_{QPSK} = Q\left(\sqrt{CNR}\right). \quad (6)$$

A comparison of equation (5) with equation (6) reveals that, for a given *CNR*, the addition of the secondary information bits in the hierarchical system causes the QPSK receivers to have a larger BER, and hence introduces a penalty to these receivers. This penalty in BER is denoted as $P_{BER}$, and is defined by the following equation

$$Q\left(\sqrt{\frac{CNR}{P_{BER}}}\right) = \frac{1}{2}Q\left(\frac{1-\lambda}{\sqrt{1+\lambda^2}}\sqrt{CNR}\right) + \frac{1}{2}Q\left(\frac{1+\lambda}{\sqrt{1+\lambda^2}}\sqrt{CNR}\right). \quad (7)$$

The BER penalty $P_{BER}$, also a function of $\lambda$ and *CNR*, represents the additional carrier power that is needed in the hierarchical system so that the QPSK receivers can have the same BER as in QPSK system without the secondary information. More precisely, in the hierarchical system with the carrier to noise ratio of *CNR,* the QPSK receivers can achieve the same raw BER as in the QPSK system with a carrier to noise ratio of $\frac{CNR}{P_{BER}}$. Plots of $P_{BER}$ for various values of $\lambda$ and *CNR* are shown in Figure 4.

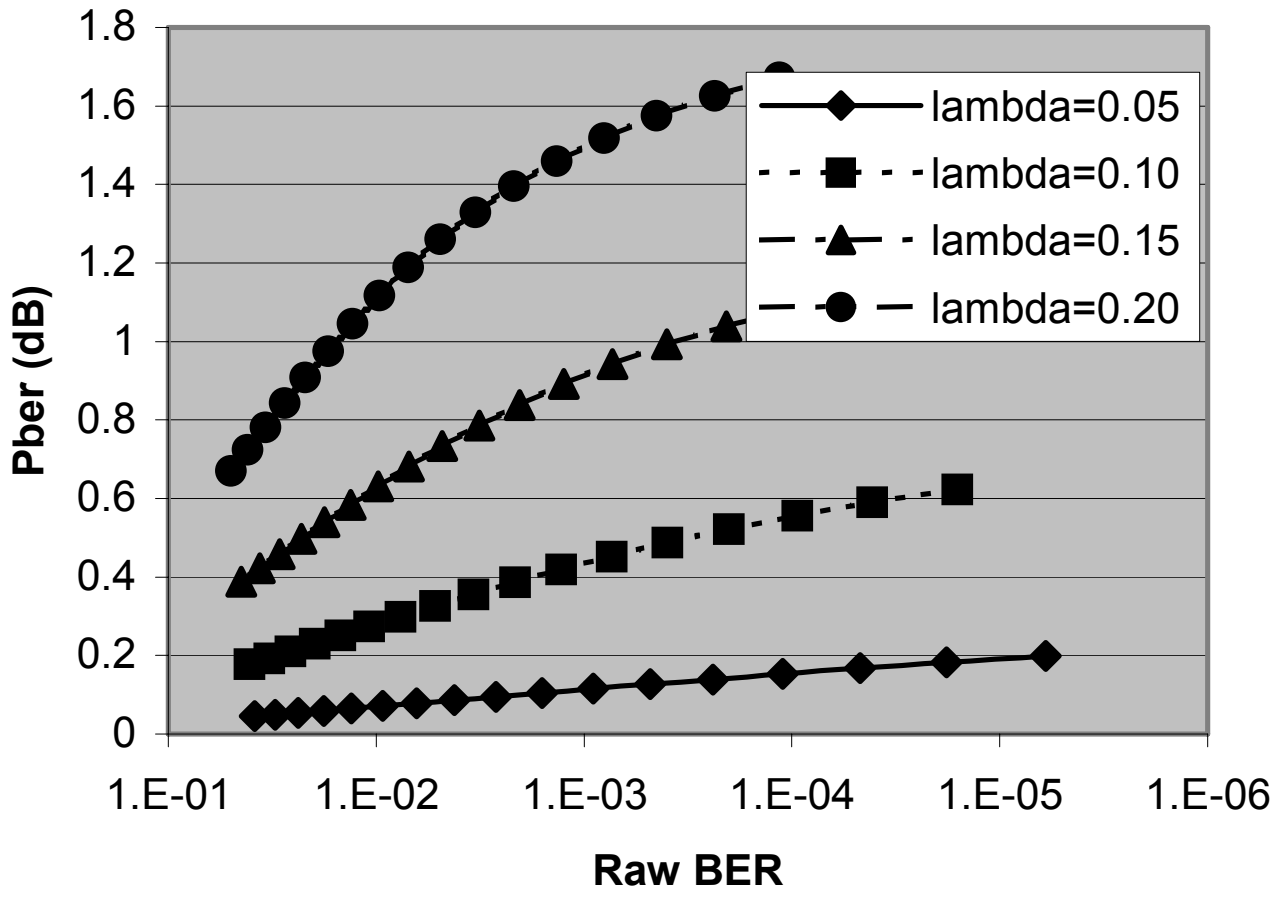

**Figure 4** Penalty in BER

As shown in Figure 4, at raw BER of 2.E-2 (corresponding to *CNR*=7dB), the penalty in BER is less than 0.25dB for $\lambda = 0.1$ and less than 0.5dB for $\lambda = 0.15$. In other words, for the QPSK receivers to achieve a raw BER of 2.E-2, a 0.25dB higher signal power is needed in the hierarchical system than the QPSK system if $\lambda = 0.1$, and a 0.5dB higher signal power is needed if $\lambda = 0.15$. The penalty is even smaller at higher BER. These results are consistent with the analysis of MNR penalty, and they again confirm that the penalty is acceptable for $\lambda \le 0.15$.

## IV. DESIGN OF HIERARCHICAL RECEIVERS

We now turn to address how to design new generation receivers to process and receive both basic and secondary information in hierarchical constellation. New generation receivers must be able to extract all bits from the 16QAM constellation of Figure 2(b). Since the basic and secondary information bits are channel encoded differently (see below), they need to be decoded separately by two decoders, and the two decoders can be used to aid each other. A block diagram of a receiver is shown in Figure 5.

After front-end demodulation, the recovered signal is sent to two separate decoders, one for the basic information bits, and the other for secondary information bits. Both decoders employ some Soft-In-Soft-Out (SISO) decoding algorithms, such as the BCJR algorithm [2, 3]. The output of the decoders, usually in the form of probability or likelihood for a bit to be a 0 or 1, is used in the decision blocks to extract the basic and information bits, respectively.

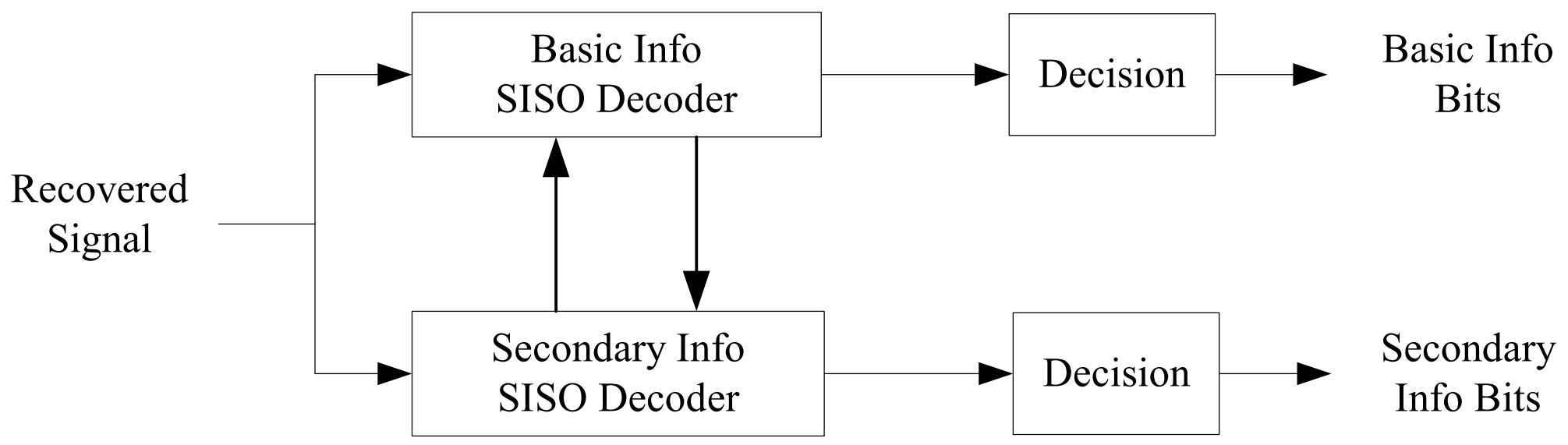

**Figure 5** New Generation Receiver Block Diagram

The channel coding for the basic information is already specified in the original QPSK system, and therefore, the decoder for the basic information bits must match the encoder of the original system because, one is not free to select a code for the basic information at the time of designing the hierarchical system. Commonly used error correction codes in deployed QPSK systems are RS-convolutional concatenated codes. Although most receivers in such systems use a Viterbi-RS decoding, which has hard-decision output, SISO type decoding algorithms exist for RS-convolutional concatenated codes, see [1, 11].

Unlike the basic information channel, one is free to choose a channel coding for the secondary information bits at the time when the hierarchical system is designed. Turbo codes [8] or other advanced codes with high coding gain may be used for the secondary information bits. These codes are well suited for an SISO type of decoding. We will not dwell on what specific decoding algorithms to use in the hierarchical receivers, because they are out of scope of this paper. Instead, we assume that an SISO decoder with *a posteriori* probability (APP) decoding [2, 3, 8] is used for each of the basic and secondary information decoders, and we propose an iterative method between the two decoders to improve the performance of both.

In an APP decoding, *a posteriori* probability $P(x=1\,|\,y)$, or $P(x=0\,|\,y)$, is estimated, where $x$ is the transmitted bit, and $y$ is the received symbol. *A posteriori* probability is the probability that a transmitted bit $x$ is 1 (or 0), given the received symbol $y$. For details of APP decoding, we refer the reader to [2, 3, 8] and the references listed therein. To perform the APP decoding, the following probabilities need to be computed:

1) The probability that the received symbol is $y$ when the transmitted bit is 0 and 1, respectively,

$$P(y\,|\,x=1) \text{ and } P(y\,|\,x=0)\,; \tag{8}$$

2) The probability that the transmitted bit is 1 (or 0)

$$P(x=1) \text{ or } P(x=0)\,. \tag{9}$$

In the following, we will present how to compute these values for the hierarchical modulation of Figure 2(b).

The mapping in Figure 2(b) allows the real and imaginary components, I and Q, of the received signal to be treated separately. Therefore, when considering probability values in equations (8) and (9), we consider the real component, I, only, as the imaginary part can be treated identically. The constellation for the real component is shown in Figure 6. In Figure 6, a "?" represents a bit carried by the imaginary component and it is irrelevant in the current discussion.

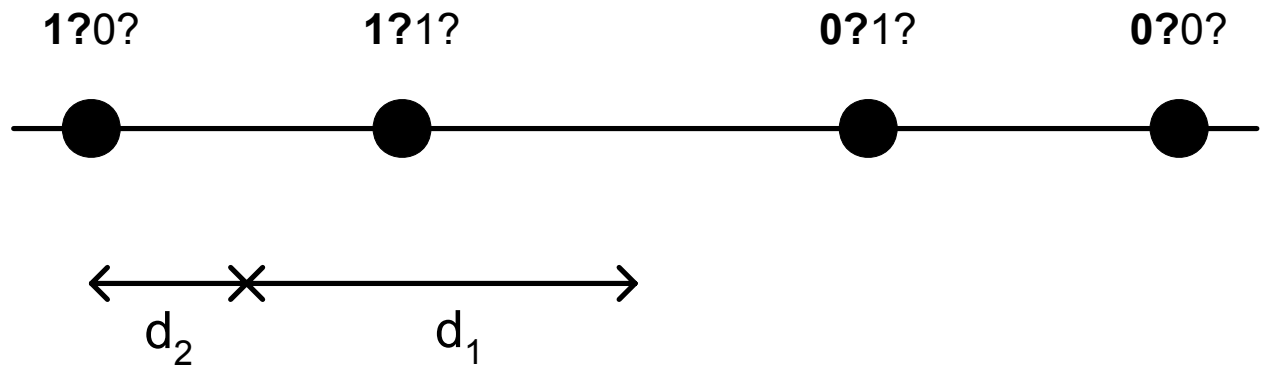

**Figure 6** One dimentional hierarchical constellation

Let $x_b$ be the transmitted basic information bit from real component, and $x_s$ be the transmitted secondary information bit from real component[1]. We will compute the probability values in equations (8) and (9) for each of $x_b$ and $x_s$. The probability that the received symbol is $y$ when the transmitted basic information bit $x_b$ is 1 is given by

[1] More precise notation should be $x_b^I$ and $x_s^I$, in order to differentiate these with bits from the imaginary component. However, the superscript $I$ is omitted since we are only considering the real component, and there is no risk of confusion.

$$P(y \mid x_b = 1) = P(x_s = 1)P(y \mid x_b = 1, x_s = 1) + P(x_s = 0)P(y \mid x_b = 1, x_s = 0). \quad (10)$$

Similarly, the probability that the received symbol is $y$ when the transmitted secondary information bit $x_s$ is 1 is given by

$$P(y \mid x_s = 1) = P(x_b = 1)P(y \mid x_b = 1, x_s = 1) + P(x_b = 0)P(y \mid x_b = 0, x_s = 1). \quad (11)$$

The conditional probability values in equations (10) and (11) are readily computed by using the probability density function of Gaussian distribution,

$$f(x) = \frac{1}{\sigma\sqrt{2\pi}} \exp\left(-\frac{x^2}{2\sigma^2}\right), \quad (12)$$

where $\sigma^2$ is the noise power. It can be shown,

$$P(y \mid x_b = 1, x_s = 1) = f(y + (1-\lambda)d_1), \quad (13)$$

because $x_b = 1, x_s = 1$ corresponds to the point labeled **1?**1? in Figure 6, and the point is located at $-(d_1 - d_2) = -(1-\lambda)d_1$. Similarly,

$$P(y \mid x_b = 1, x_s = 0) = f(y + (1+\lambda)d_1),$$
$$P(y \mid x_b = 0, x_s = 1) = f(y - (1-\lambda)d_1), \quad (14)$$
$$P(y \mid x_b = 0, x_s = 0) = f(y - (1+\lambda)d_1).$$

Now, the probability values $P(x_b = 1)$, $P(x_b = 0)$, $P(x_s = 1)$, and $P(x_s = 0)$ of equations (10) and (11) need to be computed. It is in the computation of these values that the iteration between the basic and secondary information decoders can significantly improve the performance of each decoder. The iteration between the two decoders can be carried out in the following way:

1) Initially, the decoders have no other knowledge on what is transmitted, and hence both basic and secondary information decoders start with

$$P(x_s = 1) = P(x_s = 0) = \frac{1}{2}, \quad (15)$$

$$P(x_b = 1) = P(x_b = 0) = \frac{1}{2}. \quad (16)$$

By using equations (10), (11) and (15) and (16), both decoders proceed to perform APP decoding, and at the end, the basic information decoder computes the probability values $P_b^{(0)}(x_b = 1)$ and $P_b^{(0)}(x_b = 0)$. Similarly, the secondary information decoder computes the probability values $P_s^{(0)}(x_s = 1)$ and $P_s^{(0)}(x_s = 0)$.

2) At iteration *k*, *k*=1,2,3, …, the basic information decoder performs APP decoding by using

$$P(x_s = t) = P_s^{(k-1)}(x_s = t), \quad P(x_b = t) = P_b^{(k-1)}(x_b = t) \quad t = 1,\ 0. \quad (17)$$

and equation (10). At the end of decoding, the decoder computes the values $P_b^{(k)}(x_b = 1)$ and $P_b^{(k)}(x_b = 0)$. The secondary information decoder performs APP decoding by using

$$P(x_b = t) = P_b^{(k-1)}(x_b = t), \quad P(x_s = t) = P_s^{(k-1)}(x_s = t) \quad t = 1,\ 0. \quad (18)$$

and equation (11). At the end of decoding, the decoder computes the values $P_s^{(k)}(x_s = 1)$ and $P_s^{(k)}(x_s = 0)$.

The iteration continues until a preset maximum number of iterations, *K*, is reached. The values $P_b^{(K)}(x_b = 1)$, $P_b^{(K)}(x_b = 0)$, and $P_s^{(K)}(x_s = 1)$ and $P_s^{(K)}(x_s = 0)$ are then used in the decision blocks to determine the basic and secondary information bits.

It is worthwhile to note that there is a bias in the Karnaugh map style Gray mapping of Figure 2(b). As Figure 6 and equations (13) and (14) demonstrate, it is more likely for a basic information bit to be an error bit when the secondary information bit is 1 than when the secondary information bit is 0. For example, it is more likely for the point labeled **1?**1? to be mistaken as **0?**1? or **0?**0? than the point labeled **1?**0?.

More precisely, the BER of basic information bits when the secondary information bit is 1 is given by

$$BER_B\Big|_{x_s=1} = Q\left(\frac{1-\lambda}{\sqrt{1+\lambda^2}}\sqrt{CNR}\right),$$

and the BER of basic information bits when the secondary information bit is 0 is given by

$$BER_B\Big|_{x_s=0} = Q\left(\frac{1+\lambda}{\sqrt{1+\lambda^2}}\sqrt{CNR}\right).$$

Therefore, the transmission of bit 1 in the secondary information is unfavorable to the basic information bits. If such a bias is undesirable, a mapping in Figure 7 can be used as an alternative. In the mapping of Figure 7, an 1 in the I component of the secondary information bits is more likely to cause an error in the basic information, but a 0 in the Q component of secondary information bits is more likely to cause an error. Since I, Q channels are independent and identical, 0 and 1 in the secondary information make the same contribution to the degrading of the basic information bits.

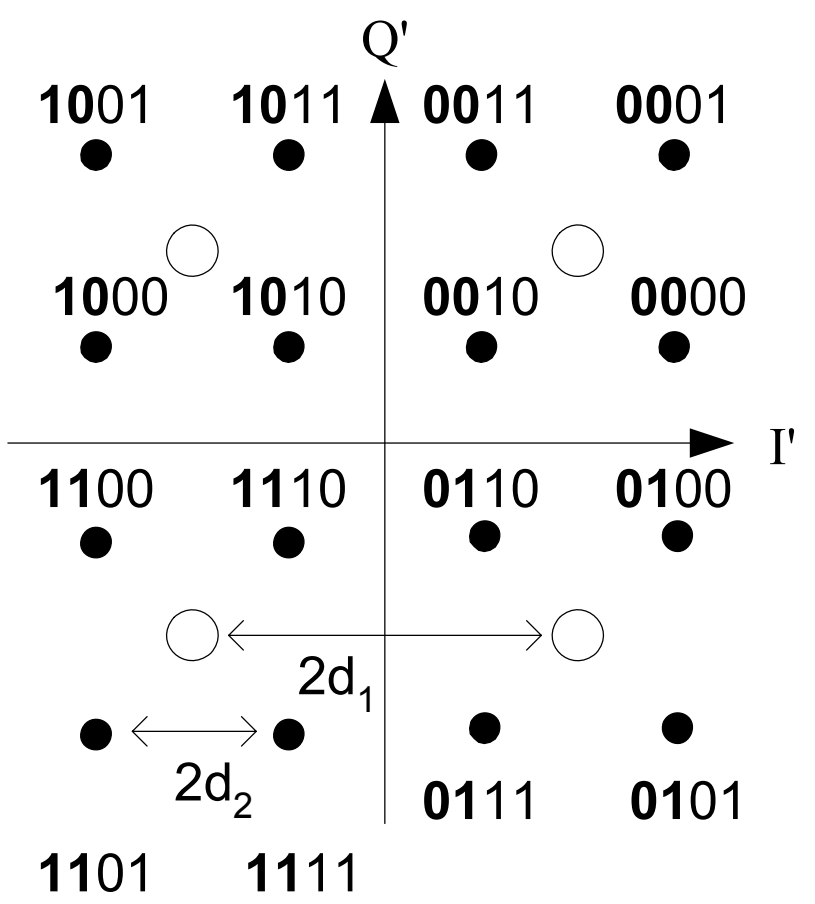

**Figure 7** Balanced hierarchical constellation

Next, we analyze the capacity of the channel that carries the secondary information bits. It can be shown that the error probability for the secondary information bits over the AWGN channel is

$$BER_S = Q\left(\frac{\lambda}{\sqrt{1+\lambda^2}}\sqrt{CNR}\right) + \frac{1}{2}Q\left(\frac{2-\lambda}{\sqrt{1+\lambda^2}}\sqrt{CNR}\right) - \frac{1}{2}Q\left(\frac{2+\lambda}{\sqrt{1+\lambda^2}}\sqrt{CNR}\right). \quad (19)$$

For small values of $\lambda$, as is needed to have a small penalty to the QPSK receivers, BER of the secondary information is much higher than that of the basic information. For the secondary information to be received with the same reliability as the basic information, the channel coding for the secondary information must have a very high coding gain, and consequently, the effective bit rate of the secondary information will be much lower than that of the basic information.

We now proceed to give a heuristic estimate of the effective bit rate of the secondary information bits by using a technique of spread spectrum. An estimate of the probability of errors for a QAM constellation is given in [9, p.338] as

$$P \approx K \cdot Q\left(\sqrt{2BT \cdot \eta_A \cdot CNR}\right) \quad . \quad (20)$$

In equation (20), *K* is some constant, *B* is the channel bandwidth, and *T* is the symbol duration and

$$\eta_A = \frac{d_{\min}^2}{4\sigma_A^2}, \quad (21)$$

where $d_{\min}$ is the minimum distance between the points of the constellation, and $\sigma_A^2 = 2(1+\lambda^2)d_1^2$ . For the basic information,

$$T = T_b,\ d_{\min} = 2(1-\lambda)d_1. \quad (22)$$

For the secondary information,

$$T = T_s,\ d_{\min} = 2\lambda\, d_1. \quad (23)$$

Substituting (22) and (23) into (20), we have the following estimates for the probabilities of errors for the basic and secondary information

$$P_b \approx K \cdot Q\left(\sqrt{BT_b \frac{(1-\lambda)^2}{1+\lambda^2} CNR}\right) \quad (24)$$

$$P_s \approx K \cdot Q\left(\sqrt{BT_s \frac{\lambda^2}{1+\lambda^2} CNR}\right) \quad (25)$$

For the secondary information bits to be received as reliably as the basic information bits, we equate the right hand sides of equations (24) and (25) to get

$$T_s = \frac{(1-\lambda)^2}{\lambda^2} T_b. \quad (26)$$

Since both basic and secondary information symbols carry the same number of bits (two bits per symbol), equation (26) gives the following relation between the effective bit rates of the basic and secondary information:

$$\text{bitrate}_s = \frac{\lambda^2}{(1-\lambda)^2}\text{bitrate}_b. \quad (27)$$

This means that for the basic and secondary information bits to be received with same BER, the bit rate of the secondary information, $\text{bitrate}_s$, needs to be a factor of $\frac{\lambda^2}{(1-\lambda)^2}$ of the basic information rate, $\text{bitrate}_b$.

From this heuristic analysis, the secondary information can be reliably received at a rate that is 1.23% of the transmission rate of the basic information if $\lambda = 0.1$. The rate of the secondary information can be increased to 3.11% of the basic information rate if $\lambda = 0.15$. For example, for an existing QPSK satellite radio system with 100 channels of music, the hierarchical system can transmit additional data in equivalence of 1.2 channels, for a penalty of 0.25dB to the QPSK receivers, and additional data in equivalence of 3 channels for a penalty of 0.5dB.

We note that these bit rates are heuristic estimates. Similar estimates can also be derived by using the channel capacity of the AWGN channel [9, p.108]

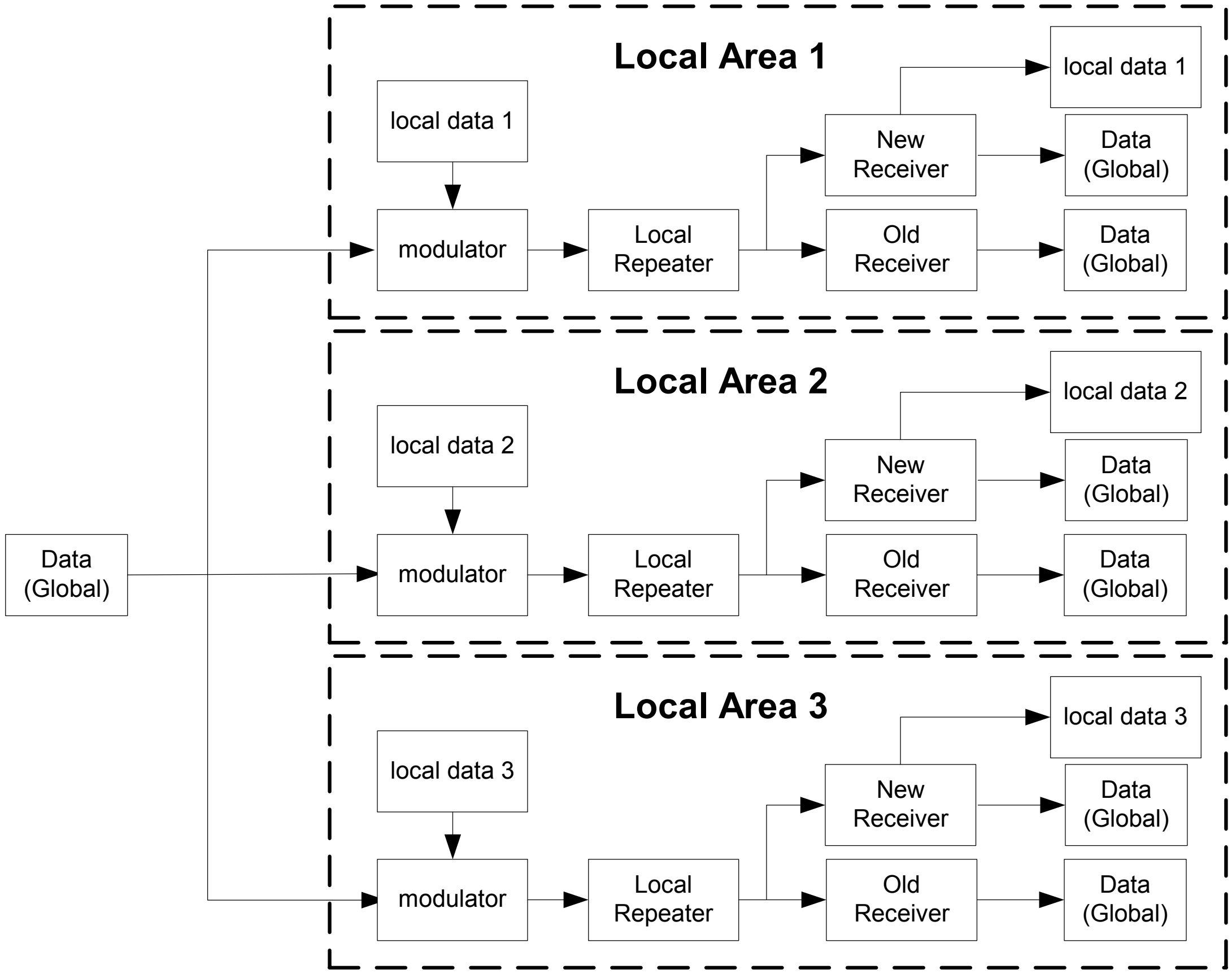

**Figure 8** Transmission of local information

$$C = \log_2(1 + SNR)\,. \qquad (28)$$

However, the actual bit rate of the secondary information depends on selections of the channel coding and decoding algorithms.

The rate of the secondary channel may be increased by using larger values for $\lambda$, at the expense of more penalties for the QPSK receivers. A recommended tradeoff would be to use a small the value of $\lambda$ when the upgraded system is initially introduced, in order to reduce the penalty to the QPSK receivers. As more new generation receivers are deployed, and the originally designed receivers are gradually phasing out, the value of $\lambda$ can be increased to raise the bit rate of the secondary information.

## V. LOCAL INFORMATION

In this section, we describe how the hierarchical modulation can be used to transmit local information efficiently in a network with local repeaters. In satellite broadcast systems such as digital satellite radio, signals are transmitted to mobile receivers. In some areas such as metropolitan regions, signals from the satellites are weak due to blockage such as tall buildings. It is a common practice to set up terrestrial repeaters in places where it is difficult to receive satellite signals. The terrestrial repeaters transmit the same data content as the satellites, but with possibly a different modulation format. With the aid of hierarchical modulation, the terrestrial repeaters can be used to provide local information that is of only interests to the local region covered by a repeater. Examples of local information include local news, traffic, weather and advertisements.

Figure 8 shows how the hierarchical modulation can be used to provide the local information efficiently in systems with a global network and local repeaters. The local information can be modulated in the secondary hierarchy, either as a part, or the entirety of the secondary information. In this way, the local information can be generated locally, and it needs not be carried in the global network, hence improving the bandwidth efficiency of the global network.

When all local repeaters transmit in a same frequency band, they form a single frequency network (SFN). Since local information content may be different from repeater to repeater, there exist interferences between neighboring repeaters. To combat the interferences, the secondary information may be modulated with orthogonal pulses such as CDMA. A small number of orthogonal codes can be predefined, so that each repeater in a cluster of neighboring repeaters, such as in a single metropolitan area, uses a different code, and no repeaters of same code will share an overlapping coverage region. The hierarchical receivers must be designed to be capable of demodulating all codes, and be able to decide which repeater's local information is used as output.

## VI. CONCLUSION

The analysis shows that the hierarchical modulation can be used effectively to upgrade a digital broadcast system in response to both the demand for higher bit rate that is made possible due to advances in technology and coding algorithm development, and the need to be backward compatible to the already deployed old receivers. The added value of the paper is the analysis of the impact that hierarchical modulation has on the already deployed receivers versus the secondary information bitrate. The hierarchical parameter $\lambda$ can be used by the system operator to control the tradeoff between the penalty to the already deployed receivers and the bit rate of the added secondary information.

This paper has been concerned with the modulation scheme. Another important issue to consider in the system upgrade is the channel coding for the secondary information, which, together with $\lambda$, will ultimately determine the actual bit rate of the secondary information. The issue of channel coding has not been addressed in this paper. Further studies are warranted to find an optimal channel coding in conjunction with the hierarchical modulation.